\documentclass[prl,twocolumn,superscriptaddress]{revtex4-1}
\usepackage[utf8]{inputenc}

\usepackage{siunitx}
\usepackage{listings}
\usepackage{amsmath}
\usepackage{amssymb}
\usepackage{graphicx}

\begin{document}

\title{Too good to be true: when overwhelming evidence fails to convince}
\author{Lachlan J.~Gunn}
\email{lachlan.gunn@adelaide.edu.au}
\affiliation{School of Electrical and Electronic Engineering, The University of Adelaide 5005, Adelaide, Australia}

\author{Fran\c{c}ois~Chapeau-Blondeau}
\email{chapeau@univ-angers.fr}
\affiliation{Laboratoire Angevin de Recherche en Ing\'enierie des Syst\`emes (LARIS), University of Angers, 62 avenue Notre Dame du Lac, 49000 Angers, France}

\author{Mark D.~McDonnell}
\email{mark.mcdonnell@unisa.edu.au}
\affiliation{School of Information Technology and Mathematical Sciences, University of South Australia, Mawson Lakes SA, 5095, Australia.}
\altaffiliation{School of Electrical and Electronic Engineering, The University of Adelaide 5005, Adelaide, Australia}

\author{Bruce R.~Davis}
\email{bruce.davis@adelaide.edu.au}

\author{Andrew Allison}
\email{andrew.allison@adelaide.edu.au}

\author{Derek Abbott}
\email{derek.abbott@adelaide.edu.au}

\affiliation{School of Electrical and Electronic Engineering, The University of Adelaide 5005, Adelaide, Australia}

\begin{abstract}

Is it possible for a large sequence of measurements or observations, 
which support a hypothesis,
to counterintuitively decrease our confidence?   Can unanimous support 
be too good to be true?
The assumption of independence is often made in good faith, however 
rarely is consideration
given to whether a systemic failure has occurred.

Taking this into account can cause certainty in a hypothesis to decrease 
as the evidence
for it becomes apparently stronger. We perform a probabilistic Bayesian analysis of 
this effect with examples based on
(i) archaeological evidence, (ii)  weighing of legal evidence,  and 
(iii) cryptographic primality testing.

We find that even with surprisingly low systemic failure rates high confidence is 
very difficult to achieve
and in particular we find that certain analyses of 
cryptographically-important numerical tests are highly
optimistic, underestimating their false-negative rate by as much as a factor of 
$2^{80}$.
\end{abstract}

\maketitle 

\section{Introduction} 

In a number of branches of science, it is now well-known that 
deleterious effects can
conspire to produce a benefit or desired positive outcome. A key example 
where this manifests
is in the field of {\it stochastic 
resonance}~\cite{benzi-stochastic-resonance,mcdonnell-stochastic-resonance,mcdonnell2009}
where a small amount of random noise can surprisingly improve system 
performance, provided some aspect of the system is nonlinear.
Another celebrated example is that of {\it Parrondo’s Paradox} where 
individually losing strategies
combine to provide  a winning outcome~\cite{harmer-parrondo,abbott-a-decade-of-parrondo}.

Loosely speaking, a small amount of `bad’ can produce a `good’ outcome.  
But is
the converse possible?  Can too much `good’ produce a `bad’ outcome?  In 
other
words, can we have too much of a good thing?

The answer is affirmative---when improvements are made that result in a 
worse
overall outcome this situation is known as {\it 
Verschlimmbesserung}~\cite{mead2005}
or {\it disimprovement}. Whilst this converse paradigm is less well 
known in the literature,
a key  example is the {\it Braess Paradox} where an attempt to improve 
traffic
flow by adding bypass routes can counterintuitively result in worse traffic
congestion~\cite{braess-braess-paradox,kameda2000, korillis1999}. Another example is the {\it 
truel}, where three
gunmen fight to the death---it turns out that under certain conditions 
the weakest gunman surprisingly
reduces his chances of survival by firing a shot at either of his 
opponents~\cite{flitney2004}.
These phenomena can be broadly  considered to fall under  the class of 
anti-Parrondo
effects~\cite{harmer2001,ethier2012}, where the inclusion of winning 
strategies fail.

In this paper, for the first time, we perform a Bayesian mathematical 
analysis to explore
the question  of multiple confirmatory measurements or observations for 
showing when
they can---surprisingly---disimprove confidence in the final outcome. We 
choose the
striking example that increasing confirmatory identifications in a 
police {\it line-up} or {\it identity parade}
can, under certain conditions, reduce our confidence that a perpetrator 
has been
correctly identified.

Imagine that as a court case drags on, witness after witness is called. 
Let us suppose thirteen
witnesses have testified to having seen the defendant commit the crime. 
Witnesses may be
notoriously unreliable, but the sheer magnitude of the testimony is 
apparently overwhelming.
Anyone can make a misidentification but intuition tells us that, with 
each additional
witness in agreement, the chance of them all being incorrect will 
approach zero. Thus one might
na\"{i}vely believe that the weight of as many as thirteen unanimous confirmations leaves us 
beyond reasonable doubt.

However, this is not necessarily the case and more confirmations can 
surprisingly disimprove our
confidence that the defendant has been correctly identified as the 
perpetrator. This type of
possibility was recognised intuitively in ancient times. Under ancient 
Jewish law~\cite{babylonian-talmud},
one could not be unanimously convicted of a capital crime---it was held 
that the absence of even
one dissenting opinion among the  judges indicated that there must 
remain some form of
undiscovered exculpatory  evidence.

Such approaches are greatly at odds with standard practice in 
engineering, where
measurements are often taken to be independent. When this is so, each 
new measurement tends
to lend support to the outcome with which it most concords. An important 
question, then, is to
distinguish between the two types of decision problem; those where 
additional measurements
truly lend support, and those for which increasingly consistent evidence 
either fails to add or
actively reduces confidence.  Otherwise, it is only later when the results come
under scrutiny that unexpectedly good results are questioned; Mendel's
plant-breeding experiments provide a good example of this~\cite{fisher-mendel,franklin-mendel},
his results matching their predicted values sufficiently well that their
authenticity has been mired in controversy since the early 20th century.

The key ingredient is the presence of a hidden failure state that 
changes the measurement
response. This change may be {\it a priori} quite rare---in the 
applications that we shall discuss,
it ranges from $10^{-1}$ to $10^{-19}$---but when several observations 
are aggregated, the
{\it a posteriori} probability of the failure state can increase 
substantially, and even come to
dominate the {\it a posteriori} estimate of the measurement response. We 
shall show that
by including error rates, this changes the information-fusion rule in a 
measurement dependent
way. Simple linear superposition no longer holds, resulting in 
non-monotonicity that leads
to these counterintuitive effects.

This paper is constructed as follows. First, we introduce an example of 
a hypothetical archaeological
find: a clay pot from the Roman era.  We consider multiple confirmatory 
measurements that decide
whether the pot was made in Britain or Italy.   Via a Bayesian analysis, 
we then show that due to
failure states,  our confidence in the pot’s origin does not improve  
for large numbers of confirmatory
measurements. We begin with this example of the pot, due to its 
simplicity and that it captures the
essential features of the problem in a clear manner.

Second, we build on this initial analysis and extend it to the problem 
of the police identity parade, showing
our confidence that a perpetrator has been identified surprisingly 
declines as the number of unanimous witnesses
becomes large. We use this mathematical framework to revisit a specific 
point of ancient Jewish law---we
show that it does indeed have a sound basis, even though it grossly 
challenges our na\"{i}ve expectation.

Third, we finish with a final example to show that our analysis has 
broader implications and can be applied to
electronic systems of interest to engineers. We chose the example of a 
cryptographic system and that
a surprisingly small bit error rate can result in a larger-than-expected 
reduction in security.

Our analyses ranging from cryptography to criminology, provide examples 
of how rare
failure modes can have a counterintuitive effect on the achievable level 
of confidence.

\section{A hypothetical Roman pot}

Let us begin with a simple scenario, the identification of the origin of a clay pot that has been dug from British soil.  Its design identifies it as being from
the Roman era, and all that remains is to determine whether it was made in Roman-occupied Britain or whether it was brought from Italy by travelling merchants.  Suppose that we are fortunate and that a test is available to distinguish between the clay from the two regions; clay from one area---let us suppose that it is Britain---contains a trace element which can be detected by laboratory tests with an error rate $p_e = 0.3$.  This is clearly excessive, and so we run the test several times.  After $k$ tests have been made on the pot, the number of errors will be binomially-distributed $E \sim \mathrm{Bin}(k, p_e)$.  If the two origins, Britain and Italy, are \emph{a priori} equally likely, then the most probable origin is the one suggested by the greatest number of samples.

Now imagine that several manufacturers of pottery deliberately introduced large quantities of this element during their production process, and that therefore it will be detected with 90\% probability in their pots, which make up $p_c = 1\%$ of those found; of these, half are of British origin.  We call $p_c$ the \emph{contamination rate}.  This is the hidden failure state to which we alluded in the introduction.  Then, after the pot tests positive several times, we will become increasingly certain that it was manufactured in Britain.  However, as more and more test results are returned from the laboratory, all positive, it will become more and more likely that the pot was manufactured with this unusual process, eventually causing the probability of British origin, given the evidence, to fall to 50\%.  This is the essential paradox of the system with hidden failure states---overwhelming evidence can itself be evidence of uncertainty, and thus be less convincing than more ambiguous data.

\subsection{Formal model}
Let us now proceed to formalise the problem above.  Suppose we have two hypotheses, $H_0$ and $H_1$, and a series of measurements $\mathbf{X} = (X_1, X_2, \ldots, X_n)$.  We define a variable $F \in \mathbb{N}$ that determines the underlying measurement distribution, $p_{X|F,H_i}(x)$.  We may then use Bayes' law to find
\begin{align}
    P[H_i | \mathbf{X}] &= \frac{P[\mathbf{X} | H_i] P[H_i]}{P[\mathbf{X}]} , \\[1.5em]
\intertext{which can be expanded by condition with respect to $F$, yielding}
                        &= \frac{\displaystyle\sum_{f} P[\mathbf{X}|H_i, f] P[H_i,F=f]}{\displaystyle\sum_{f, H_k} P[\mathbf{X}|H_k, f] P[H_k, F=f]} . \label{eqn:bayes-pot}
\end{align}

In our examples there are a number of simplifying conditions---there are only two hypotheses and two measurement distributions, reducing Eqn.~\ref{eqn:bayes-pot} to 
\begin{widetext}
\begin{align}
    P[H_i | \mathbf{X}] &= \left({1+\frac{\displaystyle\sum_{f=0}^1 P[\mathbf{X}|H_{1-i}, F=f]\; P[H_{1-i}, F=f]}{\displaystyle\sum_{f=0}^1 P[\mathbf{X}|H_i,\phantom{_1-} F=f]\; P[H_i,\phantom{_1-}F=f]}}\right)^{-1} . \label{eqn:hypothesis-posterior}
\end{align}
\end{widetext}
Computation of these \emph{a posteriori} probabilities thus requires knowledge of two distributions: the measurement distributions
$P[\mathbf{X}|H_k,F]$, and the state probabilities $P[H_i, F]$.  Having tabulated these, we may substitute them into
Eqn.~\ref{eqn:hypothesis-posterior}, yielding the \emph{a posteriori} probability for each hypothesis.
In this paper, the measurement distributions $P[\mathbf{X} | H_i, F = f]$ are all binomial, however this is
not the case in general.

\subsection{Analysis of the pot origin distribution}
In the case of the pot, the hypotheses and measurement distributions---the origin and contamination, respectively---are shown in Table~\ref{tbl:probabilities-pot}.  

\begin{table}[h!]
\begin{center}
\includegraphics[width=\linewidth]{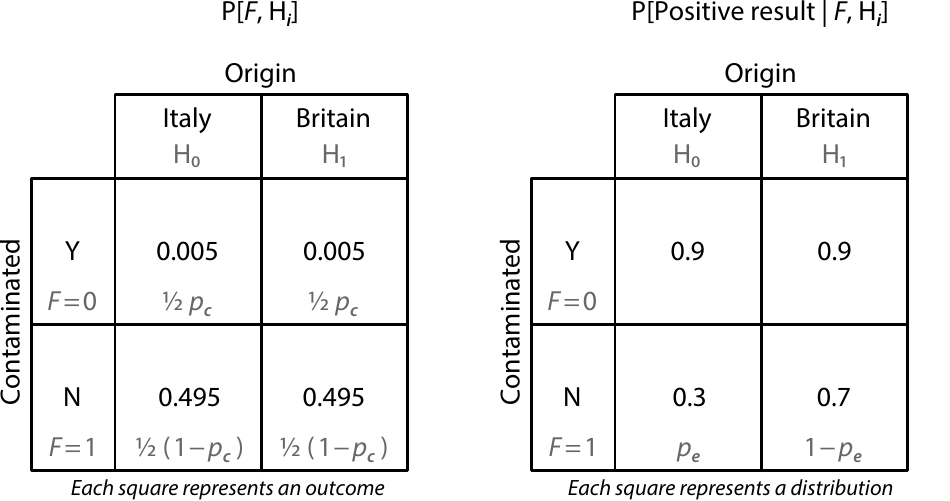}
\caption{\label{fig:pot-probabilities} The model parameters for the case of the pot for use in Eqn.~\ref{eqn:hypothesis-posterior}
with a contamination rate $p_c = 10^{-2}$.
The \emph{a priori} distribution of the origin is identically 50\% for both Britain and Italy, whether or not the pot's
manufacturing process has contaminated the results.  As a result, the two columns of $P[F, H_i]$ are identical.
The columns of the measurement distribution, shown right, differ from one another, thereby giving the test
discriminatory power.  When the pot has been contaminated, the probability of a positive result is identical
for both samples, rendering the test ineffective.%
}
\label{tbl:probabilities-pot}
\end{center}
\end{table}

Each measurement is Bernoulli-distributed, and the number of positive results is therefore
described by a Binomial distribution, with the probability mass function
\[
	P[X=x] = {n\choose x} \; p^x (1-p)^{n-x}
\]
after $N$ trials, the probability $p$ being taken from the measurement distribution
section of Table~\ref{tbl:probabilities-pot}.

\begin{figure}[h!]
\begin{center}
\includegraphics{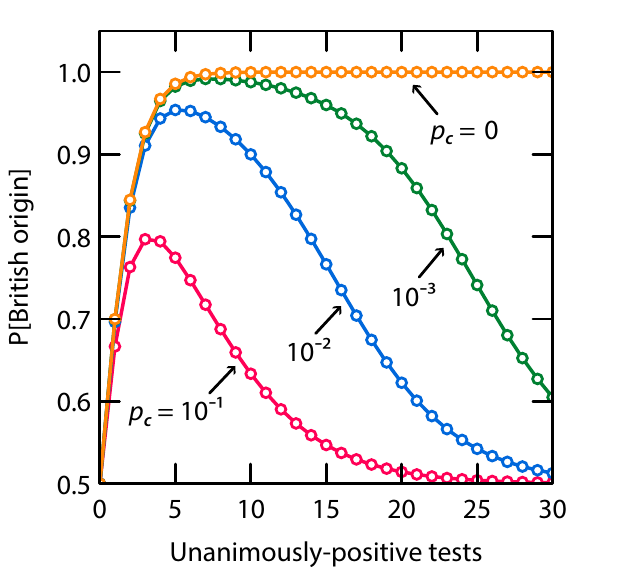}
\caption{Probability that the pot is of British origin given $n$ numbers of tests, all coming back positive,
            for a variety of contamination rates $p_c$ and a 30\% error rate.
            In the case of the pot above, with $p_c = 10^{-2}$, we see a peak at $n = 5$, after which
            the level of certainty falls back to 0.5 as it becomes more likely that
            the pot originates at a contaminating factory.  When $p_c = 0$, this is the standard Bayesian
            analysis where failure states are not considered.  We see therefore that even small contamination
            rates can have a large effect on the global behaviour of the testing methodology.%
}
\label{fig:pot-tests}
\end{center}
\end{figure}

Substituting these probability masses into Eqn.~\ref{eqn:bayes-pot}, we see
in Figure~\ref{fig:pot-tests}
that as more and more tests return positive results, we become increasingly
certain of its British heritage, but an unreasonably large
number of positive results will indicate contamination and so yield
a reduced level of certainty.

It is worth taking a moment, however, to briefly discuss the effects of
weakening certain conditions; in particular, we consider two cases:
that where the rate of contamination depends upon the origin of the pot, and
that where the results after contamination are also origin-dependant.

Where the rate of contamination depends upon the origin, evidence of contamination
provides some small evidence of where the pot came from.  Thus if 80\%
of contaminated pots are of British origin, then Figure~\ref{fig:pot-tests}
will eventually converge to $0.8$ rather than $0.5$.

If the probability of a positive test is dependent upon the origin
even when contaminated, then the behaviour of the test protocol
changes qualitatively.  Supposing that in the contaminated case
the test is substantially less accurate, the probability of British origin will
drop towards $0.5$, as now, due to the increased likelihood of contamination;
eventually, however, it will rise again towards $1.0$ as sufficient data
becomes available to make use of the less probative experiment that
we now know to be taking place.

\section{The reliability of identity parades}

We initially described the scenario of a court case, in which witness after witness
testifies to having seen the defendant commit the crime of which he is accused.
But in-court identifications are considered unreliable, and
in reality if identity is in dispute then the identification is made early in the investigation under controlled
conditions~\cite{devlin-report}.  At some point, whether before or after being charged, the suspect
has most likely been shown to each witness amongst a number of others, known as fillers, who
are not under suspicion.  Each witness is asked to identify the true perpetrator,
if present, amongst the group.

This process, known as an \emph{identity parade}
or \emph{line-up}, is an experiment intended to determine whether the
suspect is in fact the same person as the perpetrator.  It may be performed only
once, or repeated many times with many witnesses.  As human memory is
inherently uncertain, the process will include random error; if the experiment
is not properly carried out then there may also be systematic error, and this
is the problem that concerns us in this paper.

Having seen how
a unanimity of evidence can create uncertainty in the case of the unidentified pot,
we now apply the same analysis to the case of an identity parade.
If the perpetrator is not present---that is to say, if the suspect is innocent---then
in an unbiased parade the witness should be unable to choose the
suspect with a probability greater than chance.  Ideally, they would decline
to make a selection, however this does not always occur in practice~\cite{foster-eyewitness,devlin-report},
and forms part of the random error of the procedure.  If the parade is
biased---whether
intentionally or unintentionally---for example because (i) the suspect is somehow conspicuous~\cite{wogalter-distinctiveness},
(ii) the staff running the parade direct the witness towards him, (iii) by chance
he happens to resemble the perpetrator more closely than the
fillers, or (iv) because the witness holds a bias, for example
because they have previously seen the suspect~\cite{devlin-report}, then
an innocent suspect may be selected with a probability greater than chance.
This is the hidden failure state that underlies this example; we assume in
our analysis that this is completely binary---either the parade is completely
unbiased or it is highly biased against the suspect.

In recent decades, a number of experiments~\cite{malpass-lineup-instructions,foster-eyewitness}
have been carried out in order to establish the reliability of this process. Test subjects
are shown the commission of a simulated crime, whether in person
or on video, and asked to locate the perpetrator amongst a number of people.  In
some cases the perpetrator will be present, and in others not.  The former allows
estimation of the false-negative rate of the process---the rate that the witness
fails to identify the perpetrator when present---and the latter the
false-positive rate---the rate at which an innocent suspect will be mistakenly
identified.  Let us denote by $p_\mathrm{fn}$ the false-negative rate; this is equal to the proportion
of subjects who failed to correctly identify the perpetrator when he was present,
and was found in~\cite{foster-eyewitness} to be $48\%$.

Estimating the false positive rate is complicated by the fact that only one suspect is
present in the lineup---when the suspect is innocent,
an eyewitness who incorrectly identifies a filler as being
the perpetrator has correctly rejected the innocent suspect as being the perpetrator,
despite their error.  For the
purposes of our analysis, we assume that the witness selects at random in this case,
and therefore divide the 80\% perpetrator-absent selection rate
of~\cite{foster-eyewitness} by the number of participants $L=6$, yielding
a false-positive rate of $p_\mathrm{fp} = 0.133$.

Let us now suppose that there is a small probability $p_c$ that the line-up
is conducted incorrectly---for example, volunteers have been chosen who fail to
adequately match the description of the perpetrator---leading to identification of
the suspect 90\% of the time, irrespective of his guilt.  For the sake of analysis
we assume that if this occurs, it will occur for all witnesses, though in practice
the police might perform the procedure correctly for some witnesses and not others.
The probability of the
suspect being identified for each of the cases is shown in Table~\ref{tbl:probability-of-identification}.

\begin{table}
	\centering
	\includegraphics[width=\linewidth]{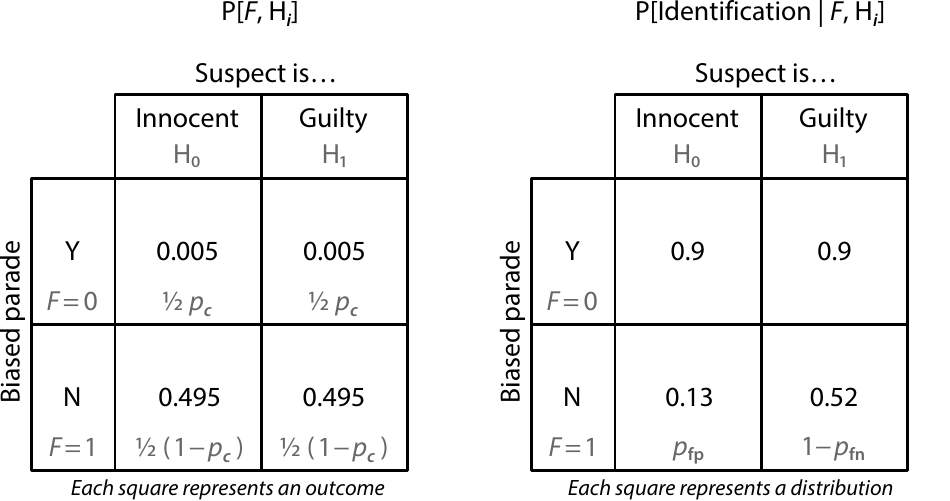}
	\caption{The model parameters for the hypothetical identity parade.
			In a similar fashion to the first example, we assume
			\emph{a priori} a 50\% probability of guilt.  In this case,
			the measurement distributions are substantially
			assymmetric with respect to innocence and guilt, unlike
			Table~\ref{tbl:probabilities-pot}.}
			
	\label{tbl:probability-of-identification}
\end{table}

\begin{figure}
\centering
\includegraphics{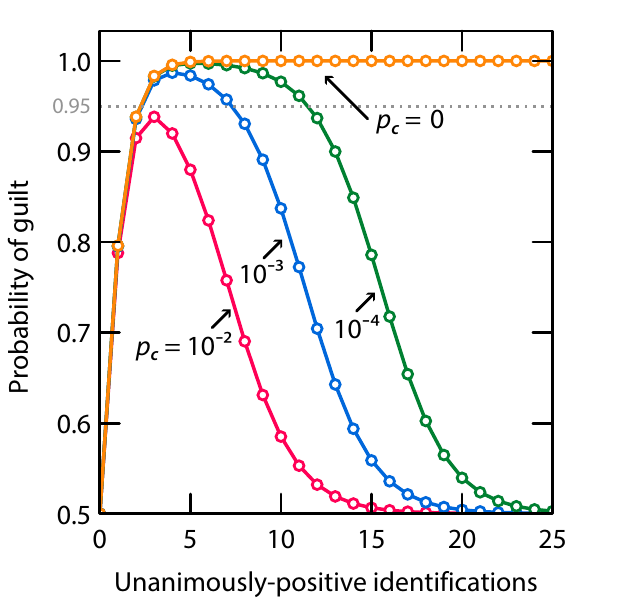}
\caption{Probability of guilt given varying numbers of unanimous line-up identifications,
            assuming a 50\% prior probability of guilt and identification accuracies
            given by~\cite{foster-eyewitness}.  Of note is that for the case that
            we have plotted here where the witnesses are unanimous, with a failure
            rate $p_c = 0.01$ it is impossible to reach $95\%$ certainty in the guilt of the
            suspect, no matter how many witnesses have been found.}
\label{fig:probability-of-guilt}
\end{figure}

If we
assume a 50\% prior probability of guilt, and independent witnesses, 
the problem is now identical to that of
identifying the pot.  The probability of guilt, given the unanimous parade
results, is shown in Figure~\ref{fig:probability-of-guilt}
as a function of the number of unanimous witnesses.

We see that after a certain number of unanimously positive
identifications the probability of guilt diminishes.  Even with only
one in ten-thousand line-ups exhibiting this bias towards the suspect,
the peak probability of guilt is reached with only five unanimous witnesses,
completely counter to intuition---in fact, with this rate of failure,
ten identifications in agreement provide less evidence of guilt than
three.  We see also that even with a 50\% prior probability of guilt,
a 1\% failure rate renders it impossible to achieve 95\% certainty
if the witnesses are unanimous.

This tendency to be biased towards a particular member of the lineup when
an error occurs has been noted~\cite[paragraph~4.31]{devlin-report}
prior to the more rigorous research
stimulated by the advent of DNA testing, leading us to suspect that
our sub-$1\%$ contamination rates are probably overly optimistic.

\section{Ancient judicial procedure}\label{sec:historical}

The acknowledgement of this type of phenomenon is not entirely new; indeed,
the adage "too good to be true" dates to the sixteenth
century~\cite[{\it good}, P5.b]{oed}.  Moreover, its influence on judicial procedure
was visible in Jewish law even in the classical era; until the Romans
ultimately removed the right of the Sanhedrin to confer death sentences,
a defendant unanimously condemned by the judges would be
acquitted~\cite[Sanhedrin 18b]{babylonian-talmud}, the Talmud stating
``If the Sanhedrin unanimously find guilty, he is acquitted.  Why? --- Because we
have learned by tradition that sentence must be postponed till the
morrow in hope of finding new points in favour of the defence''.

The value of this rule becomes apparent when we consider that the Sanhedrin
was composed, for ordinary capital offenses, of 23 members~\cite[Sanhedrin 2a]{babylonian-talmud}.  In our
line-up model, this many unanimous witnesses would indicate a probability
of guilt scarcely better than chance, suggesting that the inclusion
of this rule should have a substantial effect.

\begin{table}
	\centering
	\includegraphics[width=\linewidth]{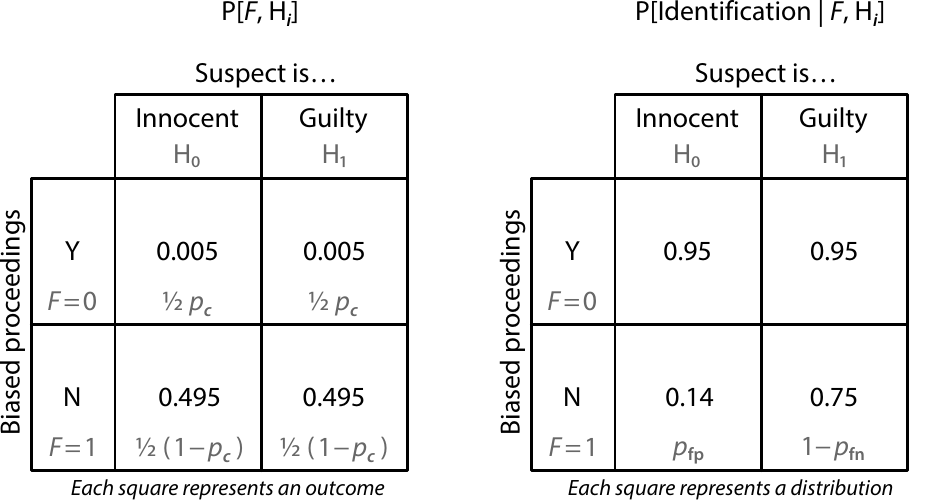}
	\caption{The model parameters for the Sanhedrin trial.
			Again, we assume an
			\emph{a priori} 50\% probability of guilt.  However, the
			measurement distributions are the results
			of~\cite[model (2)]{spencer-jury-accuracy} for juries; in contrast to
			the case of the identity parade, the false negative rate is far lower.
			Despite the trial being conducted by judges, we choose to use the jury
			results, as the judges tendancy towards conviction is not reflected
			in the highly risk-averse rabbinic legal tradition.}
			
	\label{tbl:probabilities-sanhedrin}
\end{table}

\begin{figure}
\centering
\includegraphics{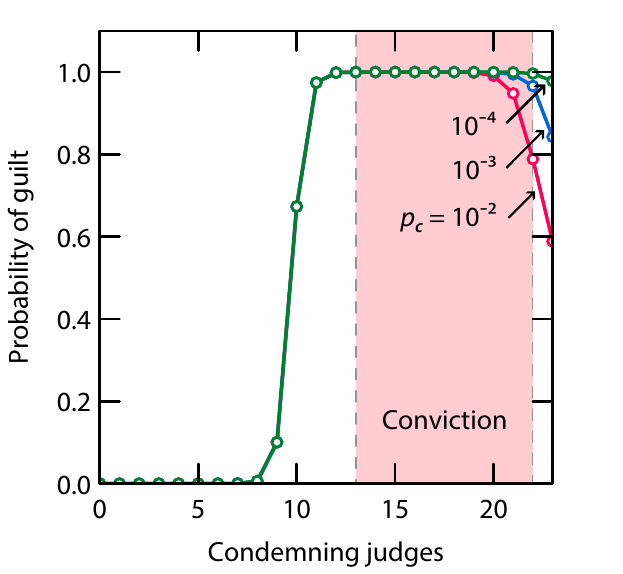}
\caption{Probability of guilt as a function of judges in agreement
            out of 23---the number used by the Sanhedrin
            for most capital crimes---for various contamination rates $p_c$.
            We assume as before that half of defendants are guilty,
            and use the estimated false-positive and false-negative 
            rates of juries from~\cite[model (2)]{spencer-jury-accuracy},
            $0.14$ and $0.25$ respectively. We arbitrarily
            assume that a `contaminated' trial will result in
            the a positive vote 95\% of the time.  The panel
            of judges numbers $23$, with conviction requiring
            a majority of two and at least one dissenting opinion~\cite[Sanhedrin]{babylonian-talmud};
            the majority of two means that the agreement of at least $13$ judges is required in order to
            to cast a sentence of death, to a maximum of $22$ votes in order to satisfy the
            requirement of a dissenting opinion.  These necessary conditions for a
	    conviction by the Sanhedrin are shown as the pink region in the graph.}

\label{fig:probability-of-guilt-sanhedrin}
\end{figure}

We show the model parameters for the Sanhedrin decision
in Table~\ref{tbl:probabilities-sanhedrin}, which we use to
compute the probability of guilt in
Figure~\ref{fig:probability-of-guilt-sanhedrin} for various
numbers of judges condemning the defendant.  We see that the
probability of guilt falls as judges approach unanimity, however
excluding unanimous decisions substantially reduces the
probability of false conviction.

It is worth stressing that the exact shapes of the curves in
Figure~\ref{fig:probability-of-guilt-sanhedrin} are unlikely
to be entirely correct; communication between the judges will
prevent their verdicts from being entirely independent, and
false-positive and false-negative rates will be very much dependent
upon the evidentiary standard required to bring charges,
the strength of the contamination when it does occur,
and the accepted burden of proof of the day.  However, it is
nonetheless of qualitative interest that with reasonable
parameters, this ancient law can be shown to have a sound
statistical basis.

\section{The reliability of cryptographic systems}

We now consider a different example, drawn from cryptography.  An
important operation in many protocols is the generation and
verification of prime numbers; the security of some protocols depends
upon the primality of a number that may be chosen by an adversary;
in this case, one may test whether it is a prime, whether by brute-force
or by using another test such as the
Rabin-Miller~\cite[p.~176]{ferguson-cryptography-engineering} test.
As the latter is probabilistic, we repeat it until we have
achieved the desired level of
security---in~\cite{ferguson-cryptography-engineering}, a
probability $2^{-128}$ of accepting a composite as prime is
considered acceptable.  However, a na\"{i}ve implementation
cannot achieve this level of security, as we will demonstrate.

The reason is that despite it being proven that each
iteration of the Rabin-Miller test will reject a composite number
with probability at least $0.75$, a real computer may fail at any time.  The chance
of this occurring is small, however it turns out that the probability
of a stray cosmic ray flipping a bit in the machine code, causing
the test to accept composite numbers, is substantially greater
than $2^{-128}$.  

\subsection{Code changes caused by memory errors}

Data provided by Google~\cite{google-dram-errors} suggests that
a given memory module has approximately an $8\%$ probability of
suffering an error in any given year, independent of
capacity.  Assuming a \SI{4}{\giga\byte} module, this results
in approximately a $\lambda=10^{-19}$ probability that any given bit
will be flipped in any given second.  We will make the
assumption that, in the machine code for the primality-testing
routine, there exists at least one bit that, if flipped,
will cause all composite numbers---or some class of composite numbers
known to the adversary---to be accepted as prime.  As an example
of how this could happen, consider the function shown in
Figure~\ref{fig:trialdivision} that implements a brute-force
factoring test.  Assuming that the input is odd, the function
will reach one of two return
statements, returning zero or one. The C compiler GCC compiles these two return statements to
\begin{lstlisting}[basicstyle=\ttfamily]
  45 0053 B8010000 movl	$1, %eax
  45      00
  46 0058 EB14     jmp	.L3
\end{lstlisting}
and
\begin{lstlisting}[basicstyle=\ttfamily]
  56 0069 B8000000 movl	$0, %eax
  56      00
\end{lstlisting}
respectively.  That is to say, it stores the return value as an
immediate into the \texttt{EAX} register and then
jumps to the cleanup section of the function, labelled
\texttt{.L3}.  The store instructions on lines $45$ and $56$
have machine-code values
\texttt{B801000000} and \texttt{B8000000000} for return
values of one and zero respectively.  These differ
by only one bit, and therefore can be transformed into
one another by a single bit-error.  If the first instruction
is turned into the second, this will cause the function to
return zero for any odd input, thus always indicating that
the input is prime.

\begin{figure}
\begin{lstlisting}[language=c,columns=flexible]
int trialdivision(long to_test)
{
	long i;
	long threshold;
	
	if(to_test % 2 == 0)
	{
		return 1;
	}
	
	threshold = (long)sqrt(to_test);
	
	for(i = 3; i <= threshold; i += 2)
	{
		if(to_test % i == 0)
		{
			return 1;
		}
	}
	
	return 0;
}
\end{lstlisting}
\caption{A function that tests for primality by attempting to
        factorise its input by brute force.}
\label{fig:trialdivision}
\end{figure}

\subsection{The effect of memory errors on confidence}

At cryptographically-interesting sizes---on the order of $2^{2000}$---roughly one in a thousand
numbers is prime~\cite[p.~173]{ferguson-cryptography-engineering}.
We might calculate the model parameters as before---for interest's sake, we have
done so in Table~\ref{tbl:probabilities-rabinmiller}---and
calculate the confidence in a number's primality after a given number of
tests.  However, this is not particularly useful, for two reasons: first,
the rejection probability of $75\%$ is a lower bound, and for randomly chosen
numbers is a substantial underestimate; second, we do not always choose numbers
at random, but rather may need to test those provided by an adversary.  In this case,
we must assume that they have tried to deceive us by providing a composite number,
and would instead like to know the probability that they will be successful.  The Bayesian
estimator in this case would provide only a tautology of the type: `given the data and the fact that
the number is composite, the number is composite'.
\begin{table}
	\centering
	\includegraphics[width=\linewidth]{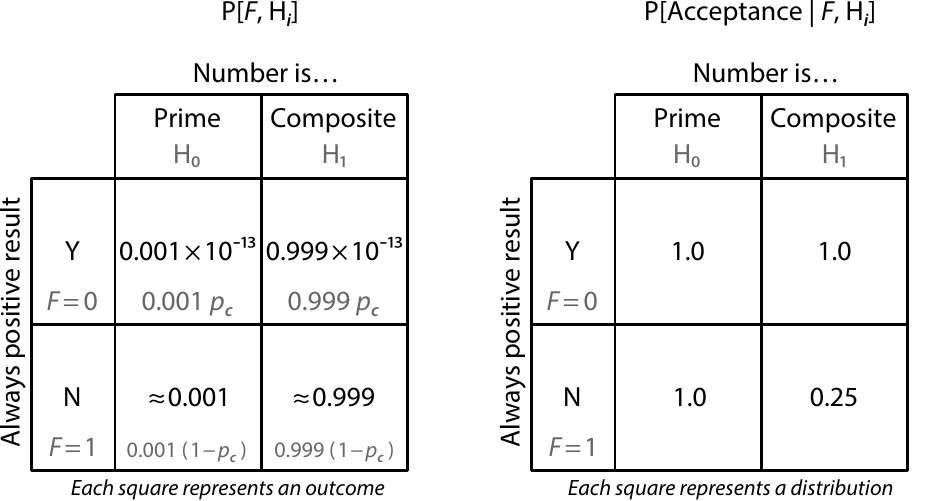}
	\caption{Model parameters for the Rabin-Miller test on random $2000$-bit numbers.
			However, we have no choice but to assume the lower bound on the
			composite-number rejection rate, and so this model is inappropriate.
			Furthermore, in an adversarial setting the attacker may intentionally choose
			a difficult-to-detect composite number, rendering the prior distribution
			optimistic.}
	\label{tbl:probabilities-rabinmiller}
\end{table}

Let us suppose that the machine containing the code is rebooted every
month, and the Rabin-Miller
code remains
in memory for the duration of this period; then, neglecting other potential
 errors that could affect the test, at the time
of the reboot the probability that the bit has flipped is now
$p_f = 2.6 \times 10^{-13}$; this event we denote $A_F$.  Let $k$ be the number of iterations performed;
the probability of accepting a composite number is at most $4^{-k}$, and we assume that the adversary
has chosen a composite number such that this is the true probability of acceptance.  We denote
the event that the prime is accepted by the correctly-operating algorithm $A_R$.  

When hardware errors are taken into account, the probability of accepting
a composite number is no longer $4^{-k}$, but
\begin{align}
	p_\mathrm{fa} &= P[A_F \cup A_R] \\
				&= P[A_F] + P[A_R] - P[A_F, A_R] . \\
\intertext{Since $A_F$ and $A_R$ are independent,}
	&= P[A_F] + P[A_R] -  P[A_F]P[A_R] \\
	&= 4^{-k} (1 - p_f) + p_f \\
	&\ge p_f .
\end{align}

\begin{figure}
\centering
\includegraphics{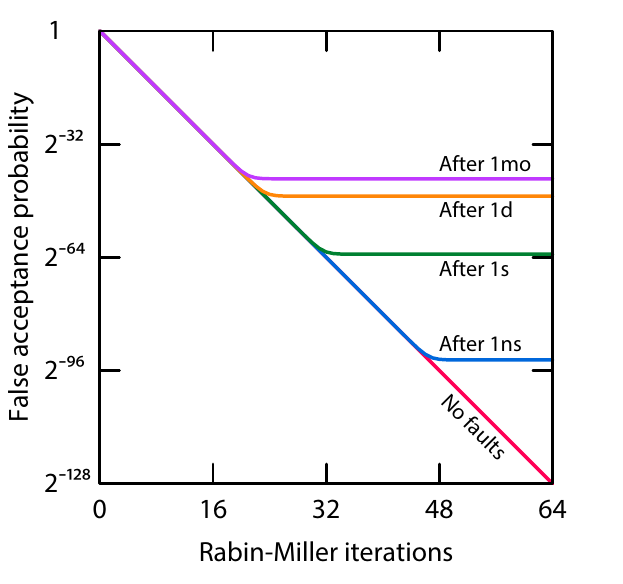}
\caption{The acceptance rate as a function of time in memory
            and the number of Rabin-Miller iterations under the
            single-error fault model described in this paper.
            An acceptance rate of $2^{-128}$ is normally chosen,
            however without error correction this cannot be
            achieved.  The false-acceptance rate after $k$ iterations
            is given by $p_\mathrm{fa}[k] = 4^{-k}(1-p_f) + p_f$, where
            $p_f$ is the probability that a fault has occurred that
            causes a false acceptance $100\%$ of the time.  We
            estimate $p_f$ to be equal to $10^{-19}T$, where $T$
            is the length of time in seconds that the code has been in
            memory.}
\label{fig:acceptance-rate}
\end{figure}

No matter how many iterations $k$ of the algorithm are performed,
this is substantially greater than the $2^{-128}$ security level that
is predicted by probabilistic analysis of the algorithm alone,
thus demonstrating that algorithmic analyses that do not take into
account the reliability of the underlying hardware can be highly
optimistic.  The false acceptance rate as a function of the number
of test iterations
and time in memory is shown in Figure~\ref{fig:acceptance-rate}.

A real cryptographic system will include many
such checks in order to make sure that an attacker has not
chosen weak values for various parameters, and a failure of
any of these may result in the system being broken,
so our calculations are somewhat optimistic.

Error-correcting-code equipped (ECC) memory will substantially
reduce the risk of this type of fault, and for regularly-accessed
regions of code---multiple times per second---will approach
the $2^{-128}$ level.  A single parity bit, as used in at least some CPU
level-one instruction caches~\cite{opteron-datasheet}, requires two bit-flips to
induce an error.  Suppose the parity is checked every
$R$ seconds, then the probability of an undetected bit-flip in
any given second is
\begin{align}
	\lambda' &= \frac{(\lambda R)^2}{R} = \lambda^2 R .
\end{align}
For code that is accessed even moderately often, this will
come much closer to $2^{-128}$.  For example, if $R = \SI{100}{\milli\second}$
then this results in a false-acceptance rate of $2^{-108}$ after one
month, much closer to the $2^{-128}$ level of security promised by
analysis of the algorithm alone.  The stronger error-correction
codes used by the higher-level caches and main memory will detect
virtually all such errors---with two-bit detection capability, the
rate of undetected bit-flips will be at most
\begin{align}
	\lambda' &= \lambda^3 R^2 ,
\end{align}
and even with check rate of only once per $\SI{100}{\milli\second}$, the rate
of memory errors is essentially zero, increasing the false-acceptance rate by a factor of
only $10^{-14}$ above the $2^{-128}$ level that would
be achieved in a perfect operating environment.

\section{Discussion}

This phenomenon is interesting in that it is commonly known
and applied heuristically, and trivial examples such as the estimation of
coin bias~\cite[section 2.1]{sivia-data-analysis} have been
well-analysed---see the appendix for a brief discussion---but these rare failure states are
rarely, if ever, considered when
a statistical approach to decision-making is applied to an entire system.  Real systems
that attempt to counter failure modes producing consistent data
tend to focus upon the detection of particular failures rather
than the mere fact of consistency.  Sometimes there is little choice---a casino
that consistently ejected gamblers on a winning streak would soon find itself
without a clientelle---however we have demonstrated
that in many cases the level of consistency needed to cast
doubt on the validity of the data is surprisingly low. 

If this is so, then we must reconsider the use of
thresholding as a decision mechanism when there is the potential for
such failure modes to exist, particularly when the consequences of
an incorrect decision are large.  When the decision rule takes the
form of a probability threshold, it is necessary to deduce an upper
threshold as well, such as was
shown in Figure~\ref{fig:probability-of-guilt-sanhedrin},
in order to avoid capturing the region indicative of a systemic
failure.

That this phenomenon was accounted for in ancient Jewish legal
practice indicates a surprising level of intuitive statistical sophistication
in this ancient law code; though predating by millennia the
statistical tools needed to perform a rigorous analysis,
our simple model of the judicial panel indicates that the
requirement of a dissenting opinion would have provided a substantial
increase in the probability of guilt required to secure a
conviction.

Applied to cryptographic systems, we see that even the minuscule
probability that one particular bit in the system's machine
code will be flipped due to a memory error over the course of a month, rendering the system
insecure, is approximately $2^{80}$ times larger than the risk
predicted by algorithmic analysis.  This demonstrates the importance
of strong error correction in modern cryptographic systems that strive
for a failure rate on the order of $2^{-128}$, a level of certainty
that appears to be otherwise unachievable without active
mitigation of the effect.

The use of naturally-occuring memory errors for DNS hijacking~\cite{dinaberg-bitsquatting}
has previously been demonstrated, and the ability of a user to disturb protected
addresses by writing to adjacent cells~\cite{kim-bitflips} has been
demonstrated, however little consideration has been given to the possibility that
this type of fault might occur simply by chance, implying that security analyses
which assume reliable hardware are substantially flawed when applied to
consumer systems lacking error-corrected memory.

We have considered only a relatively simple case, in which
there are only two levels of contamination.  However, in practical
situations we might expect any of a wide range of failure modes
varying continuously.  We have described a simple case in the
appendix, where a coin may be biased---or not---towards either
heads or tails with any strength; were one to apply this to the
case of an identity parade, for example, one would find a probability
that the suspect is indeed the perpetrator, as before, but taking into
account that there may well be slight biases that nudge the witnesses
towards or away from the suspect, not merely catastrophic ones.
The result is heavily
dependent upon the distribution of the bias, and lacking sufficient
data to produce such a model we have chosen to eschew the
complexity of the continuous approach and focus on a simple
two-level model.  An example of this approach is shown
in~\cite[p.~117]{bovens-bayesian-epistomology}.

A related concept to that which we have discussed is the Duhem-Quine
hypothesis~\cite[p.~6]{bovens-bayesian-epistomology}; this is the idea that an experiment
inherently tests hypotheses as a group---not merely the phenomenon
that we wish to examine, but also the correct function of the experimental
apparatus for example, and  that that only the desired independent
variables are being changed.  Our thesis is a related one, namely
that in practical systems the failure of these auxiliary hypotheses, though
unlikely, result in a significant reduction in confidence when
it occurs, an effect which has traditionally been ignored.

\section{Conclusion}

We have analysed the behaviour of systems that are subject
to systematic failure, and demonstrated that with relatively
low failure rates, large sample sizes are not required in order
that unanimous results start to become indicative of
systematic failure.  We have
investigated the effect of this phenomenon upon identity parades, and
shown that even with only a $1\%$ rate of failure, confidence
begins to decrease after only three unanimous identifications,
failing to reach even $95\%$.

We have also applied our analysis of the phenomenon to cryptographic systems,
investigating the effect by which confidence in the security
of a parameter fails to increase with further testing due to
potential failures of the underlying hardware.  Even with
a minuscule failure rate of $10^{-13}$ per month, this effect dominates
the analysis and is thus a significant
determining factor in the overall level of security, increasing
the probability that a maliciously-chosen parameter will be accepted
by a factor of more than $2^{80}$.

Hidden failure states such as these reduce confidence far more than
intuition leads one to believe, and must be more carefully considered
than is the case today if the lofty targets that we set for ourselves
are to be achieved in practice.





\section*{Competing interests}

We have no competing interests.

\section*{Author contributions}

LJG drafted the manuscript. LJG and DA devised the concept.
LJG, FC-B, MDM, BRD, AA, and DA carried out analyses and checking.

All authors contributed to proofing the manuscript.
All authors gave final approval for publication.



\section*{Funding}

Lachlan J.~Gunn is a Visiting Scholar at the University of Angers, France, supported
by an Endeavour Research Fellowship from the Australian Government.

Mark D.~McDonnell is supported by an Australian Research Fellowship (DP1093425) and Derek Abbott

is supported by a Future Fellowship (FT120100351), both from the Australian Research Council (ARC).
\appendix

\section{Analysis of a biased coin}

It is worth adding a brief discussion of a simple and well-known problem that
has some relation to what we have discussed, namely the question of whether
or not a coin is biased.  We follow the Bayesian approach given in~\cite{sivia-data-analysis}.

They use Bayes' law in its proportional form,
\begin{align}
	P[Q|\mathrm{\{data\}}] \propto P[\mathrm{\{data\}}|Q] P[Q] , \label{eqn:bayesprop}
\end{align}
where $Q$ is the probability that a coin-toss will yield heads.  Various
prior distributions $Q[H]$ can be chosen, a matter that we will discuss momentarily.

As the coin tosses are independent, the data can be boiled down to a
binomial random variable $X \sim \mathrm{Bin}(p,n)$, where $n$ is the number
of coin tosses made.  Substituting the binomial probability mass function into
Eqn.~\ref{eqn:bayesprop}, they find that
\begin{align}
	P[Q|X] \propto Q^X (1-Q)^{n-X} P[Q] .
\end{align}
As the number of samples $n$ increases, this becomes increasingly peaked around
the value $Q = X/n$, this `peaking' effect limited by the shape of $P[Q]$.
As the number of samples increases, the
$Q^X (1-Q)^{n-X}$ part of the expression eventually comes to dominate the shape
of the posterior distribution $P[Q|X]$, and we have no choice but to believe that the
coin genuinely does have a bias close to $X/n$.

In the examples previously discussed, we have assumed that bias is very unlikely;
in the coin example, this corresponds to a prior distribution $P[Q]$ that is strongly
clustered around $Q = 0.5$; in this case, a very large number of samples
will be necessary in order to conclusively reject the hypothesis that the coin is unbiased
or nearly so.  However, eventually this will occur, and the posterior distribution will
change; when this occurs, the system has visibly failed---a casino using the coin will
decide that they are not in fact playing the game that they had planned, and must
cease before their loss becomes catastrophic.  This is much like in the case of the
Sanhedrin---if too many judges agree, the system has failed, and should not be
considered reliable.

\bibliographystyle{ieeetran}

\begin{thebibliography}{10}
\providecommand{\url}[1]{#1}
\csname url@samestyle\endcsname
\providecommand{\newblock}{\relax}
\providecommand{\bibinfo}[2]{#2}
\providecommand{\BIBentrySTDinterwordspacing}{\spaceskip=0pt\relax}
\providecommand{\BIBentryALTinterwordstretchfactor}{4}
\providecommand{\BIBentryALTinterwordspacing}{\spaceskip=\fontdimen2\font plus
\BIBentryALTinterwordstretchfactor\fontdimen3\font minus
  \fontdimen4\font\relax}
\providecommand{\BIBforeignlanguage}[2]{{%
\expandafter\ifx\csname l@#1\endcsname\relax
\typeout{** WARNING: IEEEtran.bst: No hyphenation pattern has been}%
\typeout{** loaded for the language `#1'. Using the pattern for}%
\typeout{** the default language instead.}%
\else
\language=\csname l@#1\endcsname
\fi
#2}}
\providecommand{\BIBdecl}{\relax}
\BIBdecl

\bibitem{benzi-stochastic-resonance}
R.~Benzi, A.~Sutera, and A.~Vulpiani, ``The mechanism of stochastic
  resonance,'' \emph{Journal of Physics A: Mathematical and General}, vol.~14,
  no.~11, pp. L453--L457, 1981.

\bibitem{mcdonnell-stochastic-resonance}
M.~D. McDonnell, N.~G. Stocks, C.~E.~M. Pearce, and D.~Abbott, \emph{Stochastic
  Resonance: From Suprathreshold Stochastic Resonance to Stochastic Signal
  Quantization}.\hskip 1em plus 0.5em minus 0.4em\relax Cambridge University
  Press, 2008.

\bibitem{mcdonnell2009}
M.~D. McDonnell and D.~Abbott, ``{What is stochastic resonance? Definitions,
  misconceptions, debates, and its relevance to biology},'' \emph{PLoS
  Computational Biology}, 2009, art.~e1000348.

\bibitem{harmer-parrondo}
G.~P. Harmer and D.~Abbott, ``Game theory: losing strategies can win by
  {Parrondo's} paradox,'' \emph{Nature}, vol. 402, p. 864, 1999.

\bibitem{abbott-a-decade-of-parrondo}
D.~Abbott, ``Asymmetry and disorder: a decade of {Parrondo's} paradox,''
  \emph{Fluctuation and Noise Letters}, vol.~9, no.~1, pp. 129--156, 2010.

\bibitem{mead2005}
M.~N. Mead, ``Columbia program digs deeper into arsenic dilemma,''
  \emph{Environ.~Health Perspect.}, vol. 113, no.~6, pp. A374--A377, 2005.

\bibitem{braess-braess-paradox}
D.~Braess, ``Über ein {Paradoxon} aus der {Verkehrsplanung},''
  \emph{Unternehmensforschung}, vol.~12, pp. 258--268, 1969.

\bibitem{kameda2000}
H.~Kameda, E.~Altman, T.~Kozawa, and Y.~Hosokawa, ``Braess-like paradoxes in
  distributed computer systems,'' \emph{IEEE Transactions on Automatic
  Control}, vol.~45, no.~9, pp. 1687--1691, 2000.

\bibitem{korillis1999}
Y.~A. Korilis, A.~A. Lazar, , and A.~Orda, ``Avoiding the {Braess} paradox in
  noncooperative networks,'' \emph{Journal of Applied Probability}, vol.~36,
  no.~1, pp. 211--222, 1999.

\bibitem{flitney2004}
A.~P. Flitney and D.~Abbott, ``Quantum two- and three-person duels,''
  \emph{J.~Opt.~B: Quantum Semiclass.~Opt.}, vol.~6, no.~8, pp. S860--S866,
  2004.

\bibitem{harmer2001}
G.~P. Harmer, D.~Abbott, P.~G. Taylor, and J.~M.~R. Parrondo, ``Brownian
  ratchets and {Parrondo's} games,'' \emph{Chaos}, vol.~11, no.~3, pp.
  705--714, 2001.

\bibitem{ethier2012}
S.~N. Ethier and J.~Lee, ``Parrondo’s paradox via redistribution of wealth,''
  \emph{Electron.~J.~Probab.}, vol.~17, 2012, art.~20.

\bibitem{babylonian-talmud}
I.~Epstein, Ed., \emph{The Babylonian Talmud}.\hskip 1em plus 0.5em minus
  0.4em\relax Soncino Press, London, 1961.

\bibitem{fisher-mendel}
R.~A. Fisher, ``Has {Mendel's} work been rediscovered?'' \emph{Annals of
  Science}, vol.~1, no.~2, pp. 115--137, 1936.

\bibitem{franklin-mendel}
A.~Franklin, \emph{Ending the {Mendel-Fisher} Controversy}.\hskip 1em plus
  0.5em minus 0.4em\relax University of Pittsburgh Press, 2008.

\bibitem{devlin-report}
P.~Devlin, C.~Freeman, J.~Hutchinson, and P.~Knights, \emph{Report to the
  {Secretary of State} for the {Home Department} of the Departmental Committee
  on Evidence of Identification in Criminal Cases}.\hskip 1em plus 0.5em minus
  0.4em\relax HMSO, London, 1976.

\bibitem{foster-eyewitness}
R.~A. Foster, T.~M. Libkuman, J.~W. Schooler, and E.~F. Loftus,
  ``Consequentiality and eyewitness person identification,'' \emph{Applied
  Cognitive Psychology}, vol.~8, no.~2, pp. 107--121, 1994.

\bibitem{wogalter-distinctiveness}
M.~S. Wogalter and D.~B. Marwitz, ``Suggestiveness in photospread lineups:
  similarity induces distinctiveness,'' \emph{Applied Cognitive Psychology},
  vol.~6, no.~5, pp. 443--452, 1992.

\bibitem{malpass-lineup-instructions}
R.~S. Malpass and P.~G. Devine, ``Eyewitness identification: lineup
  instructions and the absence of the offender,'' \emph{Journal of Applied
  Psychology}, vol.~66, no.~4, pp. 482--489, 1981.

\bibitem{oed}
\emph{Oxford {English Dictionary}}.\hskip 1em plus 0.5em minus 0.4em\relax
  Oxford University Press, http://oed.com/, accessed 2015-10-06.

\bibitem{spencer-jury-accuracy}
B.~D. Spencer, ``Estimating the accuracy of jury verdicts,'' \emph{Journal of
  Empirical Legal Studies}, vol.~4, no.~2, pp. 305--329, 2007.

\bibitem{ferguson-cryptography-engineering}
N.~Ferguson, B.~Schneier, and T.~Kohno, \emph{Cryptography Engineering: Design
  Principles and Practical Applications}.\hskip 1em plus 0.5em minus
  0.4em\relax Wiley, Indianapolis, USA, 2010.

\bibitem{google-dram-errors}
B.~Schroeder, E.~Pinheiro, and W.-D. Weber, ``{DRAM} errors in the wild: A
  large-scale field study,'' in \emph{Proceedings of the Eleventh International
  Joint Conference on Measurement and Modeling of Computer Systems}, SIGMETRICS
  '09, Seattle, WA, USA, 2009, pp. 193--204.

\bibitem{opteron-datasheet}
{Advanced Micro Devices}, ``{AMD Opteron} processor product datasheet,''
  http://support.amd.com/TechDocs/23932.pdf, 2007, accessed 2015-10-12.

\bibitem{sivia-data-analysis}
D.~S. Sivia and J.~Skilling, \emph{{Data Analysis: A Bayesian Tutorial}}.\hskip
  1em plus 0.5em minus 0.4em\relax Oxford University Press, 2006.

\bibitem{dinaberg-bitsquatting}
A.~Dinaberg, ``Bitsquatting: {DNS} hijacking without exploitation,'' in
  \emph{Proceedings of BlackHat Security}, Las Vegas, USA, 2011.

\bibitem{kim-bitflips}
Y.~Kim, R.~Daly, J.~Kim, C.~Fallin, J.~H. Lee, D.~Lee, C.~Wilkerson, K.~Lai,
  and O.~Mutlu, ``Flipping bits in memory without accessing them: An
  experimental study of {DRAM} disturbance errors,'' in \emph{Proc.~IEEE 41st
  {Annual} {International} {Symposium} on {Computer} {Architecuture}},
  Minneapolis, MN, USA, 2014, pp. 361--372.

\bibitem{bovens-bayesian-epistomology}
L.~Bovens and S.~Hartmann, \emph{Bayesian Epistemology}.\hskip 1em plus 0.5em
  minus 0.4em\relax Oxford, 2004.

\end{thebibliography}

\end{document}